\documentclass[12pt]{article}
\usepackage{amsmath, amssymb}
\usepackage[latin1]{inputenc}
\usepackage{mathrsfs}
\usepackage{cite}
\usepackage{dsfont}

\begin{document}
\date{}
\title{$SU(1,1)$ solution for the Dunkl oscillator in two dimensions and its coherent states}

\author{M. Salazar-Ram\'{\i}rez$^{a}$\footnote{{\it E-mail address:} escomphysics@gmail.com.mx}, D. Ojeda-Guill\'en$^{a}$, \\ R. D. Mota$^{b}$, and
 V. D. Granados$^{c}$} \maketitle

\begin{minipage}{0.9\textwidth}

\small $^{a}$ Escuela Superior de C\'omputo, Instituto Polit\'ecnico Nacional,
Av. Juan de Dios B\'atiz esq. Av. Miguel Oth\'on de Mendiz\'abal, Col. Lindavista,
Del. Gustavo A. Madero, C.P. 07738, Ciudad de M\'exico, Mexico.\\

\small $^{b}$ Escuela Superior de Ingenier{\'i}a Mec\'anica y El\'ectrica, Unidad Culhuac\'an,
Instituto Polit\'ecnico Nacional, Av. Santa Ana No. 1000, Col. San
Francisco Culhuac\'an, Del. Coyoac\'an, C.P. 04430, Ciudad de M\'exico, Mexico.\\

\small $^{c}$ Escuela Superior de F{\'i}sica y Matem\'aticas,
Instituto Polit\'ecnico Nacional, Ed. 9, Unidad Profesional Adolfo L\'opez Mateos, Del. Gustavo A. Madero, C.P. 07738, Ciudad de M\'exico, Mexico.\\

\end{minipage}

\begin{abstract}
We study the Dunkl oscillator in two dimensions by the $su(1,1)$ algebraic method.
We apply the Schr\"odinger factorization to the radial Hamiltonian of the Dunkl oscillator
to find the $su(1,1)$ Lie algebra generators.
The energy spectrum is found by using the theory of unitary irreducible representations. By solving analytically the Schr\"odinger equation, we construct the Sturmian basis for the unitary irreducible representations of the $su(1,1)$ Lie algebra.
We construct the $SU(1,1)$ Perelomov radial coherent states for this problem and compute their time evolution.

\end{abstract}

\section{Introduction}
In Relativistic and Non Relativistic Quantum Mechanics there are few problems which can be solved exactly. Within these exactly solvable problems are those with radial symmetry, such as the harmonic oscillator, the Kepler-Coulomb and the of MICZ-Kepler problems, among others. The solution to these problems is generally obtained in two ways: analytical and algebraic. The algebraic methods provide a more elegant way to find the energy spectrum and eigenfunctions,  and generally,  these methods are related to compact and non-compact symmetries.  The $su(1,1)\sim so(2,1)$, one of the most simple Lie algebras, has been successfully applied to study many non-relativistic quantum problems, as can be seen in references \cite{ALHA1,ALHA2,ALHA3,FRANK1,FRANK2,FRANK3,WU,LEV,KERI}.

Dirac \cite{DIRLIB} and Schr\"odinger \cite{SCH1A,SCH1B,SCH1C} established the fundamental ideas of factorization since the beginnings of quantum mechanics. However, it was until the works of Infeld and Hull \cite{INF1,INF2} where the factorization methods became relevant to the study of quantum systems. They introduced  a systematic way to factorize and classify a large class of potentials. Recently, the so-called Schr\"odinger factorization has been used to construct the $su(1,1)$ Lie algebra generators and to give algebraic solutions to various physical problems \cite{DANIEL,MSAL1,MSAL2,NOS1,NOS2}.

The harmonic oscillator coherent states were introduced by Schr\"odinger \cite{SCH} in $1926$ and they are the most classical states. Glauber \cite{GLAU}, Klauder \cite{KLAU1,KLAU2} and Sudarshan \cite{SUDAR} reintroduced the concept of coherent states and  its application in quantum optics. The modern way to construct these states is by means of compact and non-compact groups and their unitary irreducible  representations. This approach for the coherent states has  been successfully applied to many problems, reported in references \cite{KLAUL,PERL,ZHANG,GAZL,KLIL}.

The Dunkl oscillator has been studied in two and three dimensions in references \cite{GEN1,GEN2,GEN3}. In these works, the authors obtained the exact solutions of the model, its superintegrability and its dynamical symmetry (called Schwinger-Dunkl algebra).
The two-dimensional solutions of the Dunkl oscillator in Cartesian coordinates are given in terms of the generalized Hermite polynomials, and in polar coordinates its angular and radial solutions  are given in terms of the Jacobi and Laguerre polynomials, respectively.

The aim of the present work is to solve the Dunkl oscillator in two dimensions by using the $su(1,1)$ Lie algebra and to construct its Perelomov radial coherent states. We apply the Schr\"odinger factorization to the radial Hamiltonian of the radial Dunkl oscillator. After generalize the factorization operators,  we construct the $su(1,1)$ Lie algebra generators, being one of the generators of the algebra proportional to the radial Hamiltonian. We construct the Sturmian basis for the unitary irreducible representations of this algebra. The energy spectrum and the radial wave functions are obtained.  With the $su(1,1)$ generators and the Sturmian basis we construct the
Perelomov coherent states and their time evolution of the two-dimensional Dunkl oscillator.

This work is organized as follows. In Section $2$ we obtain the differential equations for the radial and angular parts of the Dunkl oscillator in the plane in polar coordinates. In Section $3$, we apply the Schr\"odinger factorization method to the radial Hamiltonian to obtain the three generators of the $su(1,1)$ Lie algebra. The energy spectrum is computed by using the theory of irreducible representations. By solving analytically the Schr\"odinger radial equation, the Sturmian basis for the unitary irreducible representations of the $su(1,1)$ is obtained. In Section $4$, we construct the $SU(1,1)$ Perelomov radial  coherent states and their time evolution. Finally, we give some concluding remarks.

\section{The Dunkl oscillator in two dimensions in polar coordinates}

The Dunkl derivative $D_{x_i}^{\mu_{x_i}}$ is defined by \cite{DUNKL}
\begin{equation}
D_{x_i}^{\mu_{x_i}}=\frac{\partial}{\partial_{x_i}}+\frac{\mu_{x_i}}{x_i}\left(\mathbb{I}-R_{x_i}\right), \quad\quad x_i\in\{x,y\}.
\end{equation}
In this expression $R_{x_i}$ is the reflection operator with respect to the plane $x_i=0$, which produces the following action
\begin{equation}
R_xf(x,y)=f(-x,y), \quad\quad R_yf(x,y)=f(x,-y).
\end{equation}
With this operator, the Hamiltonian of the isotropic Dunkl oscillator in two dimensions is
\begin{equation}
H=-\frac{1}{2}\left[(D_{x}^{\mu_{x}})^2+(D_{y}^{\mu_{y}})^2\right]+\frac{1}{2}\left[x^2+y^2\right],
\end{equation}
where the term $\left(D_{x_i}^{\mu_{x_i}}\right)^2$ explicitly is given by
\begin{equation}
\left(D_{x_i}^{\mu_{x_i}}\right)^2=\frac{\partial^2}{\partial x_i^2}+2\frac{\mu_{x_i}}{x_i}\frac{\partial}{\partial x_i}- \frac{\mu_{x_i}}{x_i^2}\left[\mathbb{I}-R_{x_i}\right].
\end{equation}
In polar coordinates $x=r\cos{\varphi}$, $y=r\sin{\varphi}$ the Hamiltonian can be written as
\begin{equation}
H=A_r+\frac{1}{r^2}B_{\varphi},
\end{equation}
where $A_r$ and $B_{\varphi}$ are
\begin{equation}
A_r=-\frac{1}{2}\left[\frac{\partial^2}{\partial r^2}+\frac{1}{r}\frac{\partial}{\partial r}\right]-\frac{1}{r}(\mu_x+\mu_y)\frac{\partial}{\partial r}+\frac{1}{2}r^2,
\end{equation}
\begin{equation}
B_{\varphi}=-\frac{1}{2}\frac{\partial^2}{\partial \varphi^2}+\left(\mu_x\tan{\varphi}-\mu_y\cot{\varphi}\right)\frac{\partial}{\partial \varphi}
+\frac{\mu_x (\mathbb{I}-R_x)}{2\cos^2{\varphi}}+\frac{\mu_y (\mathbb{I}-R_y)}{2\sin^2{\varphi}}.
\end{equation}
For simplicity, we set $\mu_x$  and $\mu_y$  equal  to the constants  $\mu_1$ and $\mu_2$, respectively. If we propose that the eigenfunction $\Psi(r,\varphi)$ of the Schr\"odinger equation $H\Psi=\mathcal{E}\Psi$ is of the form $\Psi(r,\varphi)=R(r)\Phi(\varphi)$,
we obtain the differential equation for the angular part
\begin{equation}
B_{\varphi}\Phi(\varphi)-\frac{l^2}{2}\Phi(\varphi)=0,\label{angular}
\end{equation}
and the radial Hamiltonian of the problem \cite{GEN1}
\begin{equation}
H_rR(r)=A_rR(r)+\frac{l^2}{2r^2}R(r)=\mathcal{E}R(r),\label{radial}
\end{equation}
where the term $l^2/2$ is the separation constant. The solutions of the angular part (equation (\ref{angular})) depend on
the eigenvalues $s_1$, $s_2$ with $s_i=\pm1$ of the reflection operators $R_1$, $R_2$. These functions are reported in
references \cite{GEN1,GEN2} and are given by
\begin{equation}
\Phi_m^{(s_1,s_2)}(\varphi)=\eta_m \cos^{e_1}{\varphi}\sin^{e_2}{\varphi}P_{m-e_1/2-e_2/2}^{(\mu_2+e_2-1/2,\mu_1+e_1-1/2)}(\cos{2\varphi}),
\end{equation}
where $P_n^{(\alpha,\beta)}(x)$ are the classical Jacobi polynomials, $\eta_m$ is a normalization factor and
$e_1$, $e_2$ are the indicator functions for the eigenvalues of the reflections $R_1$ and $R_2$, i.e.:

\begin{equation*}
e_i=  \left\lbrace
  \begin{array}{l}
 \hspace{-0.15cm} 0, \quad\text{ if } s_i=1, \\
 \hspace{-0.15cm} 1, \quad\text{ if } s_i=-1, \\
  \end{array}
  \right.
\end{equation*}
for $i=1,2$. The constant $m$ is a positive half-integer when $s_1s_2=-1$, whereas $m$ is a
non-negative integer when $s_1s_2=1$. Moreover, for $m=0$, only the $s_1=s_2=1$ state exists. In all these parity
cases, the separation constant is
\begin{equation}
l^2=4m(m+\mu_1+\mu_2).\label{lcuad}
\end{equation}
From the orthogonality relation of the Jacobi polynomials, it can be deduced that the angular part
of the Dunkl oscillator satisfies \cite{GEN1}
\begin{equation}
\int_0^{2\pi}\Phi_m^{(s_1,s_2)}(\varphi)\Phi_{m'}^{(s'_1,s'_2)}(\varphi)|\cos{\phi}|^{2\mu_1}|\sin{\phi}|^{2\mu_2}d\phi=\delta_{m,m'}\delta_{s_1,s'_1}\delta_{s_2,s'_2},
\end{equation}
with the normalization constant $\eta_m$ defined as
\begin{equation}
\eta_m=\left[\frac{\left(2m+\mu_1+\mu_2\right)\Gamma\left(m+\mu_1+\mu_2+\frac{e_1}{2}+\frac{e_2}{2}\right)\left(m-\frac{e_1}{2}-\frac{e_2}{2}\right)!}
{2\Gamma\left(m+\mu_1+\frac{e_1}{2}-\frac{e_2}{2}+\frac{1}{2}\right)\Gamma\left(m+\mu_2+\frac{e_2}{2}-\frac{e_1}{2}+\frac{1}{2}\right)}\right]^{1/2}.
\end{equation}
The radial eigenfunctions and the energy spectrum of the Dunkl oscillator in two dimensions shall be obtained in the following Section by using an $su(1,1)$ algebraic method.

\section{The $su(1,1)$ algebraic solution of the Dunkl oscillator }

In this Section we shall use the Schr\"odinger factorization method \cite{SCH1A,DANIEL} to obtain the energy spectrum and the radial eigenfunctions of the Dunkl oscillator. In order to remove the first derivative of equation (\ref{radial}),  we propose
\begin{equation}
R(r)=\frac{U(r)}{r^{\frac{1}{2}(1+2\mu_1+2\mu_2)}}.\label{RU}
\end{equation}
Thus, the radial equation for the function $U(r)$ is
\begin{equation}
\frac{d^2U(r)}{dr^2}+\left[2\mathcal{E}-r^2-\frac{l^2-\frac{1}{4}+(\mu_1+\mu_2)^2}{r^2}\right]U(r)=0.\label{difU}
\end{equation}
The differential equation for $U(r)$ can be written in the more appropriate form
\begin{equation}
\left(-r^2\frac{d^2}{dr^2}+r^4-2\mathcal{E}r^2\right)U(r)=\left(\frac{1}{4}-l^2-(\mu_1+\mu_2)^2\right) U(r).\label{ecudi2a}
\end{equation}
In order to apply the Schr\"odinger factorization to the left-hand side of equation (\ref{ecudi2a}), we propose
\begin{equation}\label{sch}
\left(r\frac{d}{dr}+ar^2+b\right)\left(-r\frac{d}{dr}+cr^2+f\right)U(r)=gU(r),
\end{equation}
where $a$, $b$, $c$, $f$ and $g$ are constants to be determined. Expanding this
expression and comparing it with equation (\ref{ecudi2a}) we obtain
\begin{equation}
a=c=\pm1,\quad f=\mp\mathcal{E}-\frac{1}{2},\quad b=\mp\mathcal{E}-\frac{3}{2},\nonumber
\end{equation}
\begin{equation}
g=-l^2-(\mu_1+\mu_2)^2-(\mathcal{E}\pm1)^2+\frac{1}{2}.
\end{equation}
Hence, the differential equation for $U(r)$ (equation (\ref{ecudi2a})) is factorized as
\begin{align}
\left(\mathcal{J}_\mp\mp1\right)\mathcal{J}_\pm U(r)
=-\frac{1}{4}\left\{\left(\mathcal{E}\pm1\right)^2+\left[l^2+(\mu_1+\mu_2)^2-\frac{1}{2}\right]\right\}U(r),\label{Facdesc1}
\end{align}
where
\begin{equation}
\mathcal{J}_{\pm}=\frac{1}{2}\left(\mp r\frac{d}{dr}+r^2-\mathcal{E}\pm\frac{1}{2}\right),
\end{equation}
are the so-called Schr\"odinger operators. In what follows we use the standard procedure to generalize these
operators to close the $su(1,1)$ Lie algebra.  From equation (\ref{difU}) we define
\begin{equation}
B_0 U(r)\equiv\frac{1}{4}\left[-\frac{d^2}{dr^2}+r^2+\frac{l^2-\frac{1}{4}+(\mu_1+\mu_2)^2}{r^2}\right]U(r)=\frac{\mathcal{E}}{2}U(r).\label{B0}
\end{equation}
This equation allows to generalize the operators $\mathcal{J}_{\pm}$ to the second order ones
\begin{equation}\label{schop}
B_{\pm}\equiv \frac{1}{2}\left[\mp r\frac{d}{dr}+r^2-2B_0\mp\frac{1}{2}\right].
\end{equation}
By direct calculation we can show that the operators $B_{0}$ and  $B_{\pm}$ close the $su(1,1)$ Lie algebra (equation A.1 of Appendix).
We notice  that the states for the unitary irreducible representations of this Lie algebra  are  $|k,n \rangle=U(r)$.
However, if we apply the operators $B_0$ and $B_{\pm}$ on the functions $U(r)=r^{\frac{1}{2}(1+2\mu_1+2\mu_2)}R(r)$,
we get the operators
\begin{align}\label{A0}
A_0&=\frac{1}{4}\left(-\frac{d^2}{dr^2}-\frac{(1+2\mu_1+2\mu_2)}{r}\frac{d}{dr}+\frac{l^2}{r^2}+r^2\right),\\\label{A+}
A_\pm&=\frac{1}{2}\left(\pm r\frac{d}{dr}-r^2+2K_0\pm(1+\mu_1+\mu_2)\right),
\end{align}
whose unitary irreducible representations are given now  in terms of the radial functions (Sturmian basis) $|k,n \rangle=R(r)$. Notice
that these new operators also close the $su(1,1)$ Lie algebra and that the radial Hamiltonian $H_r$ (equation (\ref{radial})) and the third generator $A_0$ satisfies $A_0=\frac{1}{2}H_r$.

The action of the Casimir operator $A^2$ (see Appendix) on the radial function $R(r)$ is
\begin{equation}
A^2R(r)=\frac{1}{4}\left[(\mu_1+\mu_2)^2+l^2-1\right]R(r).\label{caspart}
\end{equation}
By substituting equation (\ref{lcuad}) into equation (\ref{caspart}) and from the theory of unitary irreducible representation it follows that
\begin{equation}
\frac{1}{4}\left[(\mu_1+\mu_2)^2+4m(m+\mu_1+\mu_2)-1\right]R(r)=k(k-1)R(r).
\end{equation}
Thus, the group number $k$ (Bargmann index) is
\begin{equation}
k=m+\frac{1}{2}\left(\mu_1+\mu_2+1\right), \quad\quad k=-m-\frac{1}{2}\left(\mu_1+\mu_2-1\right), \label{k}
\end{equation}
and the other group number is just the radial quantum number,  $n=n_r$. From equation (\ref{radial}) and equation (\ref{k0n}) of Appendix
we obtain
\begin{equation}
H_rR(r)=2A_0R(r)=\mathcal{E}R(r) =2\left[n_r+m+\frac{1}{2}\left(1+\mu_1+\mu_2\right)\right]R(r).
\end{equation}
Therefore, the energy spectrum of the Dunkl oscillator in two dimensions is
\begin{equation}
\mathcal{E}=2(n_r+m)+\mu_1+\mu_2+1.\label{esp}
\end{equation}

The radial function $U(r)$ can be obtained from the general differential equation \cite{LEB}
\begin{equation}
u''+\left[4n+2\alpha+2-x^2+\frac{\frac{1}{4}-\alpha^2}{x^2}\right]u=0,\label{difgen}
\end{equation}
which has the particular solution
\begin{equation}
u=N_ne^{-x^2/2}x^{\alpha+1/2}L_n^{\alpha}(x^2).
\end{equation}
By substituting the energy spectrum of equation (\ref{esp}) into the differential equation
for the radial function $U(r)$ (\ref{difU}), and comparing it with equation (\ref{difgen}), we obtain the relationship
\begin{equation}
\alpha^2=l^2+(\mu_1+\mu_2)^2=4m(m+\mu_1+\mu_2)+(\mu_1+\mu_2)^2,
\end{equation}
where we have used the expression (\ref{lcuad}). Therefore, the function $U(r)$ explicitly is
\begin{equation}
U(r)=\frac{2\Gamma(n_r+1)}{\Gamma(n+2m+\mu_1+\mu_2+1)}e^{-r^2/2}r^{2m+\mu_1+\mu_2+\frac{1}{2}}L_{n_r}^{2m+\mu_1+\mu_2}(r^2).
\end{equation}
where the normalization coefficient $N_n$ was computed from the orthogonality of the Laguerre polynomials
\begin{equation}
\int_0^{\infty}e^{-x}x^{\alpha}\left[L_{n}^{\alpha}(x)\right]^2dx=\frac{\Gamma(n+\alpha+1)}{n!}.
\end{equation}
The Sturmian basis for the irreducible unitary representation of the $su(1,1)$ Lie algebra in terms
of the group numbers $n,k$ for the Dunkl oscillator in two dimensions  is
\begin{equation}
R_{n_r,m}(r)=\left[\frac{2\Gamma\left(n_r+1\right)}{\Gamma\left(n_r+2m+\mu_1+\mu_2+1\right)}\right]^{1/2}r^{2m} e^{-r^2/2}L_{n_r}^{2m+\mu_1+\mu_2}\left(r^2\right).\label{Rfin}
\end{equation}
or
\begin{equation}
R_{n,k}(r)=\left[\frac{2\Gamma(n+1)}{\Gamma\left(n+2k\right)}\right]^{1/2}r^{2k-(\mu_1+\mu_2+1)} e^{-r^2/2}L_{n}^{2k-1}\left(r^2\right). \label{st}
\end{equation}
As it was expected, it can be easily seen that for $\mu_1=\mu_2=0$ the radial functions of the two-dimensional isotropic harmonic oscillator are recovered \cite{GUR,CCG}.

\section{$SU(1,1)$ radial coherent states and their time-evolution}

In this Section we shall construct the radial coherent states for the radial function by using the Sturmian basis $R_{k,n}(r)$.
The $SU(1,1)$ Perelomov coherent states are defined as \cite{PERL}
\begin{equation}
|\zeta\rangle=D(\xi)|k,0\rangle=(1-|\xi|^2)^k\sum_{n=0}^\infty\sqrt{\frac{\Gamma(n+2k)}{n!\Gamma(2k)}}\xi^n|k,n\rangle.
\end{equation}
with $D(\xi)$ the displacement operator and $|k,0\rangle$ the lowest normalized state.
Thus, if we apply the operator $D(\xi)$ to the ground state of the Dunkl oscillator radial the functions ($n=0$ in equation (\ref{st})),
we obtain
\begin{equation}\label{est1}
R(r,\xi)=\left[\frac{2(1-|\xi|^2)^{2k}}{\Gamma(2k)}\right]r^{2k-(\mu_1+\mu_2+1)} e^{-r^2/2}\sum_{n=0}^\infty\xi^nL_{n}^{2k-1}\left(r^2\right).
\end{equation}
This sum is computed from the Laguerre polynomials generating function
\begin{equation}
\sum_{n=0}^\infty L_n^\nu(x)y^n=\frac{e^{-xy/(1-y)}}{(1-y)^{\nu+1}},
\end{equation}
resulting the closed form
\begin{equation}
R(r,\xi)=\left[\frac{2(1-|\xi|^2)^{2k}}{\Gamma(2k)(1-\xi)^{4k}}\right]r^{2k-(\mu_1+\mu_2+1)} e^{\frac{r^2}{2}\left(\frac{\xi+1}{\xi-1}\right)}.
\end{equation}
Therefore the $SU(1,1)$ radial coherent states for the Dunkl oscillator in two dimensions in terms of the physical quantum
numbers $m,\mu_1,\mu_2$ are
\begin{equation}
R(r,\xi)=\left[\frac{2(1-|\xi|^2)^{2m+(\mu_1+\mu_2+1)}}{\Gamma(2m+(\mu_1+\mu_2+1))(1-\xi)^{2(2m+(\mu_1+\mu_2+1))}}\right]r^{2m} e^{\frac{r^2}{2}\left(\frac{\xi+1}{\xi-1}\right)}.\label{cstates}
\end{equation}
The radial coherent states for the two-dimensional Dirac-Moshinsky oscillator coupled to an external magnetic field were obtained in relativistic quantum mechanics in reference \cite{NOS3}.

The time evolution of these coherent states can be easily computed since $A_0=\frac{1}{2}H_r$. Thus, the time evolution operator for our problem is
\begin{equation}\label{UTEM}
\mathcal{U}(\tau)=e^{-iH_{r}\tau/\hbar}=e^{-i\gamma 2A_0\tau/\hbar},
\end{equation}
where $\tau$ is considered as a fictitious time \cite{GUR,CCG}. Thus, the time evolution of the Perelomov coherent states is given by \cite{NOS1}
\begin{equation}\label{PERET}
|\zeta(\tau)\rangle =\mathcal{U}(\tau)|\zeta\rangle=\mathcal{U}(\tau)D(\xi)\mathcal{U}^\dag(\tau)\mathcal{U}(\tau)|k,0\rangle.
\end{equation}

From equation (\ref{k0n}) of Appendix, the time evolution of the state $|k,0\rangle$ is
\begin{equation}\label{evest1}
\mathcal{U}(\tau)|k,0\rangle=e^{-2ik\tau/\hbar}|k,0\rangle.
 \end{equation}
The similarity transformation $\mathcal{U}(\tau)D(\xi)\mathcal{U}^\dag(\tau)$ can be computed from the time evolution of the raising and lowering operators $\mathcal{U}^\dag(\tau)A_{\pm}\mathcal{U}(\tau)$.
From the BCH (Baker-Campbell-Hausdorff) formula and equations (\ref{UTEM}) and (\ref{com}) we have
\begin{align}
A_+(\tau)=&\mathcal{U}^\dag(\tau)A_+\mathcal{U}(\tau)=A_+e^{2i \tau/\hbar},\\
A_-(\tau)=&\mathcal{U}^\dag(\tau)A_-\mathcal{U}(\tau)=A_-e^{-2i \tau/\hbar}.
\end{align}
Thus, $\mathcal{U}(\tau)D(\xi)\mathcal{U}^\dag(\tau)$ can be expressed as
\begin{equation}\label{opd1}
\mathcal{U}(\tau)D(\xi)\mathcal{U}^\dag(\tau)=e^{\xi A_+(-\tau)-\xi^*A_-(-\tau)}=e^{\xi(-\tau)A_+ - \xi(-\tau)^*A_-},
\end{equation}
where $\xi(t)=\xi e^{2i\tau/\hbar}$. In this sense, if we define $\zeta(t)=\zeta e^{2i\tau/\hbar}$, the time evolution of the displacement operator in its normal form is given by
\begin{equation}\label{opedes}
D(\xi(t))=e^{\zeta(t) A_+}e^{\eta A_0}e^{-\zeta(t)^*A_-}.
\end{equation}
from equations (\ref{evest1}) and (\ref{opedes}), we obtain that the time dependent Perelomov coherent state is
\begin{equation}
|\zeta(t)\rangle=e^{-2ik\tau/\hbar}e^{\zeta(-\tau)A_+}e^{\eta A_0}e^{-\zeta(-\tau)^*A_-}|k,0\rangle.
\end{equation}
With these results, we obtain that the time evolution of the coherent state for the Dunkl oscillator $R(r,\xi,\tau)$ in the configuration space is

\begin{align}\nonumber
R\left(r,\xi(\tau)\right)=&\left[\frac{2(1-|\xi|^2)^{2m+(\mu_1+\mu_2+1)}}{\Gamma(2m+(\mu_1+\mu_2+1))(1-\xi e^{2i\tau/\hbar})^{2(2m+(\mu_1+\mu_2+1))}}\right]^{1/2}\\
\times & e^{-2i\left(m+\frac{1}{2}\left(\mu_1+\mu_2+1\right)\right)\tau/\hbar} r^{2m} e^{\frac{r^2}{2}\left(\frac{\xi e^{2i\tau/\hbar}+1}{\xi e^{2i\tau/\hbar}-1}\right)}.\label{tecs}
\end{align}

Hence, the fact that the third generator of the Lie algebra  is proportional to the radial Hamiltonian $H_r$, simplifies
the calculation of the time evolution of the coherent states. We emphasize that equations (\ref{cstates}) and (\ref{tecs}) are two of the main results of the present work.

\renewcommand{\theequation}{A.\arabic{equation}}
\setcounter{equation}{0}
\section*{Appendix: The $SU(1,1)$ Group and its coherent states}

The $su(1,1)$ Lie algebra is spanned by the generators $K_{+}$, $K_{-}$
and $K_{0}$, which satisfy the commutation relations \cite{VOU}
\begin{eqnarray}
[K_{0},K_{\pm}]=\pm K_{\pm},\quad\quad [K_{-},K_{+}]=2K_{0}.\label{com}
\end{eqnarray}
The action of these operators on the Sturmian basis $\{|k,n\rangle, n=0,1,2,...\}$ is
\begin{equation}
K_{+}|k,n\rangle=\sqrt{(n+1)(2k+n)}|k,n+1\rangle,\label{k+n}
\end{equation}
\begin{equation}
K_{-}|k,n\rangle=\sqrt{n(2k+n-1)}|k,n-1\rangle,\label{k-n}
\end{equation}
\begin{equation}
K_{0}|k,n\rangle=(k+n)|k,n\rangle,\label{k0n}
\end{equation}
where $|k,0\rangle$ is the lowest normalized state. The Casimir
operator for any irreducible representation satisfies
\begin{equation}
K^{2}=-K_{+}K_{-}+K_{0}(K_{0}-1)=k(k-1).\label{cas}
\end{equation}
The theory of unitary irreducible representations of the $su(1,1)$ Lie algebra has been
studied in several works \cite{ADAMS} and it is based on equations (\ref{k+n})-(\ref{cas}).
Thus, a representation of $su(1,1)$ algebra is completely determined by the number $k$. For the purpose of the
present work we will restrict to the discrete series only, for which
$k>0$.

The $SU(1,1)$ Perelomov coherent states $|\zeta\rangle$ are
defined as \cite{PERL}
\begin{equation}
|\zeta\rangle=D(\xi)|k,0\rangle,\label{defPCS}
\end{equation}
where $D(\xi)=\exp(\xi K_{+}-\xi^{*}K_{-})$ is the displacement
operator and $\xi$ is a complex number. From the properties
$K^{\dag}_{+}=K_{-}$ and $K^{\dag}_{-}=K_{+}$ it can be shown that
the displacement operator possesses the property
\begin{equation}
D^{\dag}(\xi)=\exp(\xi^{*} K_{-}-\xi K_{+})=D(-\xi),
\end{equation}
and the so-called normal form of the displacement operator is given by
\begin{equation}
D(\xi)=\exp(\zeta K_{+})\exp(\eta K_{0})\exp(-\zeta^*
K_{-})\label{normal},
\end{equation}
where $\xi=-\frac{1}{2}\tau e^{-i\varphi}$, $\zeta=-\tanh
(\frac{1}{2}\tau)e^{-i\varphi}$ and $\eta=-2\ln \cosh
|\xi|=\ln(1-|\zeta|^2)$ \cite{GER}. By using this normal form of the displacement
operator and equations (\ref{k+n})-(\ref{k0n}), the Perelomov coherent states are found to
be \cite{PERL}
\begin{equation}
|\zeta\rangle=(1-|\xi|^2)^k\sum_{n=0}^\infty\sqrt{\frac{\Gamma(n+2k)}{n!\Gamma(2k)}}\xi^n|k,n\rangle.\label{PCN}
\end{equation}

\section{Concluding remarks}

We solved the Dunkl oscillator in polar coordinates by using the $su(1,1)$ Lie algebra and the theory of unitary irreducible
representations. We showed that the radial Hamiltonian  of this problem possesses the $su(1,1)$ symmetry. The Lie algebra generators
were constructed from the Schr\"odinger factorization method.
The radial Sturmian basis was obtained by solving analytically the radial Schr\"odinger differential equation.
We used the Sturmian basis and  the $su(1,1)$ Lie algebra to calculate the Perelomov radial coherent states in a closed form.
Also, we computed the time evolution of these coherent states.

Our procedure can be applied successfully to  other problems, as for example the Dunkl-Coulomb problem, which is a forthcoming work.

\section*{Acknowledgments}
This work was partially supported by SNI-M\'exico, COFAA-IPN,
EDI-IPN, EDD-IPN, SIP-IPN project number $20161727$.

\end{document}